\documentclass[12pt]{article}
\usepackage{amsfonts,amsmath,amssymb}
\usepackage{scalerel}[2016/12/29]
\usepackage{graphicx}
\usepackage{tikz}

\def\be{\begin{equation}}
\def\ee{\end{equation}}
\def\bea{\begin{eqnarray}}
\def\eea{\end{eqnarray}}

\def\half{{\textstyle {1\over 2}}}

\usepackage{amsmath}
\usepackage{epsfig,graphicx}
\usepackage{graphicx}
\usepackage{tikz}

\def\be{\begin{equation}}
\def\ee{\end{equation}}
\def\bea{\begin{eqnarray}}
\def\eea{\end{eqnarray}}

\def\r{{\rho}}

\def\At{{\tilde A}}
\def\Bt{{\tilde B}}

\def\n2{{n \over 2}}

\def\rd{{\dot q}}
\def\pd{{\dot p}}
\def\rd{{\dot \rho}}

\def\d{{\partial}}
\def\pb{{\bar p}}

\definecolor{mygreen}{rgb}{0.2, 0.8, 0.2}
\definecolor{myblue}{rgb}{0.3, 0.3, 1.0}
\definecolor{myred}{rgb}{0.9, 0.2, 0.0}
\definecolor{myorange}{rgb}{0.85, 0.45, 0.0}

\begin{document}

\thispagestyle{plain}

\title{\bf Solitons in fluctuating hydrodynamics of diffusive processes}

\author{Alexios P. Polychronakos}


\maketitle

%

{\em
{\centerline{Physics Department, the City College of New York, NY 10031, USA}}
\vskip .2 cm \centerline{and}
\vskip .2 cm
{\centerline{The Graduate Center, CUNY, New York, NY 10016, USA
}}}

\begin{abstract}

We demonstrate that fluid mechanical systems arising from large fluctuations of one-dimensional statistical processes generically
exhibit solitons and nonlinear waves. We derive the explicit form of these solutions and examine their properties for
the specific cases of the Kipnis-Marchioro-Presutti model (KMP) and the
Symmetric Exclusion Process (SEP). We show that the two fluid systems are related by a nonlinear transformation but
still have markedly different properties. In particular, the KMP fluid
has a nontrivial sound wave spectrum exhibiting birefringence, while sound waves for the SEP fluid are essentially trivial.
The appearance of sound waves and soliton configurations in the KMP model is related to the onset of instabilities.

\end{abstract}

\noindent

\vskip 0.5cm

{\centerline {\it apolychronakos@ccny.cuny.edu}}
\vskip 1cm

\vfill
\eject

\tableofcontents

\section{Introduction}

Statistical systems of particles moving randomly on the line with an exclusion rule restricting their relative positions \cite{Spitz}, also known as single-file processes, constitute examples of non-equilibrium statistical processes, such as driven diffusive systems, and have been the object of intense study (for comprehensive expositions see \cite{Lig}-\cite{KRB}; for a recent concise review see \cite{Kirone}). 

An interesting aspect of these models is that in the diffusive scaling limit, where they are subject to the macroscopic fluctuation
theory \cite{MFT,MFTL,MFTR},
they give rise to a class of fluid dynamics equations. Specifically, in the scaling limit these systems satisfy a large-deviation principle, where  the rate function for rare large fluctuations can be obtained as the solution of a variational problem
\cite{Bert,Lasi}. The optimal paths for this variational problem, constrained on the fluctuations, satisfy a set of coupled
hydrodynamic equations with Hamiltonian structure \cite{Bert,KMS}. The microscopic structure of the model is encoded
in the transport coefficients that appear in the equations, giving rise to different classes of fluid dynamics \cite{HKPS}. Related results were derived in \cite{LPS,RoVa,MeSa}.

In general, the resulting fluid equations are not analytically solvable. Nevertheless, they constitute interesting and novel
fluid dynamical systems. In determining their properties a crucial question is whether they admit any solitary wave solutions.
The existence of such solutions is usually a signal, though not a guarantee, that the underlying systems are integrable.
However, at face value, these fluid dynamical systems do not appear to be integrable, with the exception of the relatively
trivial case of independent random walkers.

In this paper we study solitary wave solutions, as well as nonlinear waves and sound waves, for a class
of exclusion process fluid mechanical systems. We cast the equations of motion to a form closely related to ordinary
hydrodynamics and analyze their symmetries and constant-profile solutions. The methods used are general, but we focus
on the specific cases of the
Symmetric Exclusion Process (SEP) and the Kipnis-Marchioro-Presutti model (KMP) \cite{KMP}. We demonstrate the existence of both
solitons and nonlinear waves, derive explicit expressions for the soliton solutions and examine the dispersion relation of sound waves. Although the
SEP and KMP systems are
related, their solutions do not map one-to-one and they have markedly different properties. In particular, the KMP fluid
has a nontrivial sound wave spectrum exhibiting birefringence, while sound waves for the SEP fluid are trivial and can exist only
over an empty ($\rho=0$) or completely full ($\rho =1$) background.

Traveling wave solutions for the KMP system were obtained in \cite{ZaMe}, and our results for this system are very close to the
ones in that reference. Our main contribution consists of a general method with which soliton and wave solutions can be found for any
diffusivity and conductivity functions $D(\rho)$ and $\sigma(\rho)$, a recasting of the equations into a form closely related to dispersive hydrodynamics
and a nontrivial mapping between the SEP and KMP models.
We also comment on the implications of the solutions for the stability of the underlying diffusive systems.

\section{General formulation and symmetries}

\subsection{Review of the hydrodynamics of macroscopic fluctuations}

We present a brief introduction to Macroscopic Fluctuation Theory (MFT) and the emergence of a hydrodynamic
description of exclusion processes. This is meant to provide context and a general background to the hydrodynamic systems
studied in the sequel. Readers interested in the specifics of MFT are advised to study the original papers
\cite{Bert,Lasi} (and \cite{Derr}, for a nice physical derivation),
while experts in the field or those only interested in the hydrodynamics can skip ahead to section 2.2.

Statistical processes in one dimension are generically described in terms of the occupation number of lattice cites $i$
at a set of discrete times. In exclusion processes, these occupation numbers are 0 or 1, and they change from discrete time
$\tau$ to $\tau+1$ according to prescribed transition probabilities. In Markovian processes these probabilities depend
on the occupation numbers at time $\tau$ as well as a set of external driving parameters (``fields''). For driving fields
constant in time and a finite number of sites such processes are expected to reach an equilibrium distribution.

Due to the probabilistic nature of their time evolution, occupation numbers, whether in or out of equilibrium, are subject
to statistical fluctuations. The determination of the statistics of these fluctuations is, in general, a hard problem.
Substantial progress in this direction can be achieved in the scaling limit of a large number of sites and a large number
of time steps \cite{Bert}. In that limit, the lattice cite number $i$ and the discrete time $\tau$ are traded for
continuous variables $x,t$
\be
x = \epsilon\, i ~,~~~ t = \delta \, \tau
\ee
with $\epsilon$ and $\delta$ the (small) space and time discrete scales. Occupation numbers are traded for a macroscopic
(coarse-grained) occupation density $\rho (x)$
\be
\rho(x) = \epsilon\, {N[i, i+\Delta x/\epsilon] \over \Delta x}
\ee
with $N[i,j]$ the total occupation number of sites $i, i+1,\dots,j$ and $\Delta x \gg \epsilon$ but macroscopically small.
The parameter $L = \epsilon^{-1}$ plays the role of the Avogadro number and is macroscopically large.

In the scaling limit the average density $\rho (x,t)$ evolves as a diffusion process obeying the continuity equation
\be
{\dot \rho} = -\partial j
\label{primcon}
\ee
where overdot is $\partial_t$ and $\partial$ stands for $\partial_x$. The current $j(x,t)$ is determined by
$\rho$ itself and an external driving field $E(x,t)$ as
\be
j = -D(\rho) \partial \rho + \sigma(\rho) E
\label{constit}\ee
$D(\rho)$ is a density-dependent diffusion constant (diffusivity) while $\sigma (\rho)$ is a density-dependent conductivity.
(\ref{constit}) is a constitutive relation for the process, with the functions $D(\rho)$ and $\sigma(\rho)$ determined by
the specifics of the diffusion process.

The basic tenet of MFT is the transition probability formula for a finite excursion of the macroscopic system from a
density profile $\rho_o (x)$ at $t=t_o$ to a profile $\rho_f (x)$ at $t=t_f$. Specifically, the probability that the density
profile will follow the path $\rho(x,t)$ given that, at time $t=t_o$, $\rho = \rho_o (x)$ is given by \cite{Bert}
\be
I_{[t_o , t_f]} \left[ \rho(x,t) \right] \sim \exp\left(-\epsilon^{-1} \int_{t_o}^{t_f} dt \int dx {(j + D \partial \rho)^2 \over 2 \sigma}\right)
\label{trans}\ee
where the current $j$ is related to $\r$ through the continuity relation (\ref{primcon}) ${\dot \rho} + \partial j =0$.
This is analogous to the Einstein relation for the probability of density fluctuations in thermal equilibrium.
The basic difference is that, since it is a transition probability, it applies to systems both in and out of equilibrium.

The probability that the system will have density $\rho_f (x)$ at time $t=t_f$ given that it has density $\rho_o (x)$
at time $t=t_o$ will be given by summing the transition probability over all intermediate configurations; that is, by
the path integral of the transition amplitude (\ref{trans}) over all $\rho(x,t)$ and $j(x,t)$
satisfying (\ref{primcon}) and the boundary conditions $\rho(x,t_{o,f} ) = \rho_{o,f} (x)$. This is analogous to
Euclidean quantum field theory, with $\epsilon$ playing the role of the Planck constant. In the macroscopic limit
$\epsilon \to 0$ only the classical paths will be relevant for the probability, that is, evolutions $\rho(x,t)$ and
$j(x,t)$ that minimize the exponent in (\ref{trans}). This leads to an action principle, minimizing
\be
S = \int  dt dx \left[ {(j + D \partial \rho)^2 \over 2 \sigma} + p \left( {\dot \rho} + \partial j \right) \right]
\ee
In the above, $p(x,t)$ is a Lagrange multiplier field implementing the continuity equation constraint
${\dot \rho} + \partial j =0$. The field $j(x,t)$ is not dynamical and appears quadratically in the action,
and integrating it out (that is, substituting the solution of its equation of motion $j = \sigma \partial p - D\partial \rho$)
we obtain
\be
S = \int dt dx \left[ p {\dot \rho} + D \partial p \partial \rho - \half \sigma (\partial p)^2 \right]
\label{act}\ee
where we dropped a total $x$-derivative in the integrand.

The action (\ref{act}) determines the optimal paths $\rho(x,t)$ and also appears in the exponential of the transition
amplitude (\ref{trans}) (the Lagrange multiplier term vanishes on shell, and the total derivative contribution vanishes
for periodic spaces). Therefore, it determines the statistical distribution of $\rho(x,t)$ through its Hamilton-Jacobi
equation. Note that it involves the fundamental variable $\rho$ as well as the conjugate variable $p$. The equations of
motion in $\r$ and $p$ are first-order in time and their solution is determined by initial conditions $\rho(x,t_o), p(x,t_o)$ 
{\it or}
by boundary conditions $\rho(x,t_o), \rho(x,t_f)$, the latter being relevant to the calculation of the probability of large
fluctuations through (\ref{trans}).

The dynamical system defined by the action (\ref{act}) for $\rho,p$ has the generic form of a fluid mechanical system.
Its properties are the main object of study in this paper and are analyzed in the subsequent sections.

\subsection{A symmetric form and time reversal}

The action (\ref{act}) is of Hamiltonian form with Lagrangian density
\be
L = p \rd +D(\r)\, \partial p \, \partial \r - \half \sigma(\r) \left( \d p \right)^2
\label{Lag}\ee
$\r$ and $p$ are canonically conjugate fields and obey the equations of motion
\bea
\rd &=& -\d \left (\sigma \d p - D\d \r \right) \nonumber \\
\pd &=& -D \,\d^2 p - \half \sigma' (\d p )^2 \label{eom}
\eea
where prime stands for $\r$-derivative. We see that $\d p$ plays the role of the field $E$ in (\ref{constit}) and represents
a driving force field that creates an additional drift current $\sigma \d p$. In the minimization process it obeys its own evolution equation (\ref{eom}). For $p=0$ we recover the diffusion equation for particles in the exclusion process \cite{MFTL,KOV}, as
expected, with a diffusion current $-D \d \r$.

The Lagrangian (\ref{Lag}) in invariant under constant shifts of $p \to p+c$, leading to the conservation of the total
number of particles. It also has a time reversal invariance, and we will recast the system in terms of variables that make the
time reversal symmetry explicit. To that end, we trade the variable $\r$ for the new variable $\pb$
\be
\pb = -p + f (\r)
\ee
with the function $f(\r)$ satisfying
\be
f' = {2D \over \sigma}
\ee
{($f(\r)$ is essentially the $\r$-derivative of the free energy of the fluid.)}
The Lagrangian in terms of the variables $p$ and $\pb$, dropping total time derivatives, rewrites as
\be
L = -\r (p,\pb)\, {\pd} + \half {\sigma(p,\pb)}\, {\d p \,\d\pb}
\ee 
or, noting that $\rho=f^{-1} (p+\pb)$ depends only on $p+\pb$,
\be 
L = \half Q(p+\pb) ({\dot \pb} - {\pd}) + \half \Sigma(p+\pb) \, \d p \,\d\pb
\ee
with $Q(\,\cdot\, ) = f^{-1} ( \,\cdot\,  )$ and $\Sigma (\,\cdot\,  ) = \sigma (f^{-1} (\,\cdot\, ))$.

The above Lagrangian is manifestly invariant under the time reversal symmetry
\be
p \leftrightarrow \pb ~,~~~  t \to -t
\ee
In terms of the original variables, this means that
\be
\r_{_T} = \r (-t) ~,~~~ p_{_T} = -p(-t) + f(q(-t))
\ee
are also solutions of the equations of motion (\ref{eom}). That is, the change $p \to \pb$ can drive $\r$
to evolve backwards in time.

For independent random walkers (which we will refer to as independent particles), 
$D=1$ and $\sigma = 2\r$, so $f = \ln \r$ and the Lagrangian becomes
\be
L = \half e^{p+\pb}\, ({\dot \pb} - \pd ) + e^{p+\pb} \, \d p \, \d \pb
\ee
or
\be
L = \Psi {\dot \Phi} + \d\Psi \, \d \Phi ~,~~ \Psi = e^p ~,~~ \Phi = e^\pb
\ee
which corresponds to two decoupled diffusion and antidiffusion processes. The transition from $\r,p$ to
$\Phi,\Psi$ is the Hopf-Cole canonical transformation \cite{DG,KMS}.

For the SEP process, $D_{\rm SEP} = 1$ and $\sigma_{\rm SEP} = 2\rho (1-\r)$, so
\be
\pb = -p + \ln{\r \over 1-\r }
\ee
The SEP process has the additional particle-hole reflection symmetry
\be
\r \to 1-\r ~,~~~ p \to -p
\label{refl}
\ee
that leaves the Lagrangian and the action invariant.

For the KMP process, $D_{\rm KMP} = 1$ and $\sigma_{\rm KMP} = 2\r^2$, so
\be
\pb = -p - {1 \over \r}
\ee
Note that if $-1 < \r p < 0$ then both $p$ and $\pb$ are negative.
The KMP process has the additional scaling symmetry
\be
p \to \lambda p ~,~~ \pb \to \lambda \pb
\label{scale}\ee
that leaves the Lagrangian and the action invariant and leads to the additional conserved charge
$D = \int dx \r p$ through the conservation equation
\be
\partial_t (\r p ) + \partial_x \left(\r \d p - p \d \r + 2\r^2 p \d p \right) = 0
\ee
Unlike the free particle case, no decoupling or simplification of the equations of motion arises in the SEP and KMP cases
by any obvious change of variables.

\subsection{Mapping SEP to KMP and an additional symmetry}

Assume $\r_{1} , p_{1}$ are SEP variables with Lagrangian
\be
L = p_{1} \rd_{1} + \d p_{1} \d \r_{1} - \r_1 ( 1-\r_1 ) (\d p_1 )^2
\ee
The transformation
\be
\r_2 = \r_1 e^{-p_1} ~,~~~ p_2 = - e^{p_1}
\label{mapKS}\ee
maps the Lagrangian, up to the total time derivative term $\partial_t (\r_1 p_1 - \r_1 )$, to
\be
L = - \left[\, p_2 \rd_2 + \d p_2 \d \r_2 - \r_2^2 (\d p_2 )^2 \right]
\ee
This is the negative of the Lagrangian of the KMP process, so $\r_2 , p_2$ obey KMP equations of motion.
Note that (\ref{mapKS}) is a Hopf-Cole type canonical transformation but with a crucial additional minus sign.
The inverse transformation, mapping KMP to SEP, is
\be
\r_1 = -\r_2 p_2 ~,~~~ p_1 = \ln (- p_2 )
\label{mapSK}\ee
Any solution of the SEP process provides a solution of the KMP process. The opposite, however, is not true, as
only KMP processes with $p_2 <0$ map to acceptable SEP processes. (An alternative transformation that allows
$p_2 >0$ maps to SEP with density outside of the acceptable range $0<\r_1 <1$.) Further, the KMP fluids that
map to SEP satisfy
\be
\r_2 p_2 = - \r_1 ~~~\Rightarrow ~~~ -1 <\r_2 p_2 <0
\label{range}
\ee
which means that both $p_2$ and $\pb_2$ are negative, preventing a time reversal transformation from producing
solutions with $p_2 >0$.

The mapping (\ref{mapKS}) can be used to relate the transition probabilities for large fluctuations of the SEP model
to those of the KMP model.  The actions under the mapping, taking into account the total derivative, map as
\be
S_2 = - S_1 + \left. \r_1 ( p_1 - 1 ) \right|_{t_o}^{t_f}
\label{S12}\ee
Therefore, a large fluctuation of the SEP model maps to a corresponding large fluctuation of the KMP model with
probability related through equations (\ref{trans}) and (\ref{S12}).

The SEP-KMP mapping allows us to relate the symmetries of the models. The reflection symmetry (\ref{refl}) of the SEP
process maps into a corresponding symmetry of the KMP process:
\be
\r_2 \to - p_2 (1+\r_2 p_2 ) ~,~~~ p_2 \to {1\over p_2}
\label{Ksym}\ee
This is physical, preserving the positivity of $\r_2$, only in the range (\ref{range}), as expected. The charge symmetry
$p_1 \to p_1 + c$ of the SEP model maps to the scaling symmetry (\ref{scale}) of the KMP model. The charge
symmetry $p_2 \to p_2 + c$ of the KMP model, on the other hand, reveals a corresponding symmetry for the SEP model
\be
\r_1 \to \r_1 + c \,\r_1 e^{-p_1} ~,~~~ p_1 \to p_1 + \ln \left(1+c \,e^{-p_1} \right)
\ee
leading to the conservation of the additional charge $C = \int dx \r_1 e^{-p_1}$ through the conservation equation
\be
\partial_t \left( \r_1 e^{-p_1} \right) + \partial_x \left[ e^{-p_1} \left( \r_1 (1-2\r_1 ) \d p_1 - \d \r_1 \right) \right] = 0
\ee

We stress that the SEP-KMP mapping  presented here is distinct from a corresponding mapping identified
in \cite{DG}. In our language, the mapping in \cite{DG} applies to systems with $D(\r)=1$ and quadratic diffusion function
$\sigma (\r) = 2 A \r (B-\r)$ and consists of the linear rescaling
\be
\r \to B \r ~,~~~ p \to B^{-1} p ~~~\Rightarrow ~~~ \sigma \to 2\r(1-\r)
\label{DGtr}\ee
which leaves the Lagrangian invariant and maps the system to the SEP fluid. Our transformation, however, is nonlinear, it mixes
$\r$ and $p$ and maps the Lagrangian to {\it minus} itself. Moreover, the linear transformation (\ref{DGtr}) becomes singular
for the KMP fluid, which requires $B \to 0 , A \to -1$ and makes the mapping between solutions of the two systems
in general singular.

The significance of the existence of the SEP-KMP mapping (\ref{mapSK}) is an interesting and presumably unexplored
feature of the SEP and KMP systems. In particular, it raises the obvious question
of whether it is a member of a more general family of transformations. It would also be very useful to have a microscopic explanation of this mapping at the level of the diffusion processes. At this point we have no answer to these
questions. Given that the above two systems are, in fact, quite special from the hydrodynamical point of view, as will be
demonstrated in the next section, it is not inconceivable that a mapping exists only and specifically for these systems.

\section{Mapping to regular fluids}

The density $\r$ and conjugate variable $p$ do not directly map to regular fluid variables. We would like to express the
problem in terms of variables and equations as closely related to regular fluids as possible. To this end, we define
\be
\theta = {p - \pb \over 2}  = p - \half f
\ee
In terms of $\rho, \theta$ the Lagrangian becomes
\be
L = \theta \rd - \half \sigma ( \d\theta )^2 + {D^2 \over 2\sigma} ( \d \rho )^2
\label{fl}\ee
This has the form of a standard fluid action. Defining the fluid velocity as
\be
v = \frac{\raisebox{-1pt}{$\hskip 0.05cm\sigma$}}{\raisebox{2pt}{$\rho$}} \, \d \theta
\ee
we arrive, after some calculation and rearrangements, at the equations of motion
\bea
&&\hskip 3cm{\dot \rho} + \d (\rho v )=0 \nonumber\\
&&{\dot v}={\sigma \over 2\rho}{\Bigl({\rho^2 \over{\raisebox{2pt}{$\sigma$}}}\Bigr)''} \, v^2 \hskip 0.05cm \d\rho 
+ \sigma^2 \left({\rho \over \sigma^2} \right)' \hskip -0.05cm v \hskip 0.05cm \d v 
- \frac{\raisebox{-1pt}{$\hskip 0.05cm\sigma$}}{\raisebox{2pt}{$\rho$}} \,
\d \left[{D \over \sqrt{\sigma}} \d \left({D \over \sqrt{\sigma}}\d\rho\right)\right]
\label{fleq}\eea
The first equation is the kinematical continuity equation for the fluid, while the second is the dynamical (Newton's) equation.
The Newton equation has the general form of dispersive hydrodynamics. We note,
however, that the transport term $v \d v$ has a nonstandard $\rho$-dependent coefficient,
and there is a nonstandard dynamical $v^2 \d \rho$ term with a $\rho$-dependent coefficient. The form of both these
coefficients depends on $\sigma(\rho)$, while $D(\rho)$ enters only the interaction potential in the last term in
(\ref{fl}) and (\ref{fleq}).

Obtaining standard hydrodynamics requires further restrictions on $\sigma(\rho)$:

a) For the $v^2 \d \rho$ term to vanish
\be
\left({\rho^2 \over \sigma}\right)' = {\rm constant} \equiv a ~~~ \Rightarrow ~~~ \sigma = {\rho^2 \over a \rho + b}
\ee

b) For the transport term $v \d v$ to have a constant coefficient
\be
\sigma^2 \left({\rho \over \sigma^2}\right)' = 1 - 2\;{\partial \ln \sigma \over \partial \ln \rho} = {\rm constant}
\equiv 1-2B
~~~ \Rightarrow ~~~ \sigma = A \,\rho^B
\ee
We observe that the special cases $\sigma = 2\rho$ (independent particles) and $\sigma = 2 \rho^2$
(KMP process) satisfy both conditions. By the mapping of the previous section, the KMP fluid also
generates solutions for the SEP fluid. Therefore, we will eventually focus on the above two special cases.

The case of independent particles, $D=1$, $\sigma=2\rho$, can further be reduced to two independent 
linear processes by putting, as in section 2,
\be
\Phi = \sqrt{\rho}\, e^{-\theta} ~,~~~ \Psi = \sqrt{\rho}\, e^{\theta} ~~~\Rightarrow ~~~ L = \Phi {\dot \Psi} + \d\Phi \, \d\Psi
\ee
The equations of motion are
\be
{\dot \Phi} + \d^2 \Phi = 0 ~,~~~~  {\dot \Psi} - \d^2 \Psi =0
\ee
so $\Psi$ and $\Phi$ are a diffusing and an anti-diffusing density, respectively. (The existence of the latter is
compatible with the time reversal symmetry of the system derived in the previous section.) In this case there is also
a Galilean boost invariance
\be
\Phi (x,t ) \to e^{-{u\over 2} x + {u^2 \over 4} t} \; \Phi (x-ut,t) ~,~~~\Psi (x,t ) \to e^{ {u\over 2} x -{u^2 \over 4} t}
\; \Psi (x-ut,t)
\ee
leaving the Lagrangian and the equations of motion invariant.

This is an essentially trivial case, leaving the KMP process and the related SEP process as the main nontrivial fluids
of interest. Note that for the KMP fluid the kinematical
transport term is $-3 v \d v$, with a coefficient different from the standard value $-1$ which obtains for independent
particles, so this is a strongly nonclassical fluid.

\section{Soliton and wave configurations}

We now proceed to examine motions corresponding to a constant fluid profile traveling at speed
$u$, potentially with an underlying particle current. Such configurations are nonlinear waves, if they repeat
periodically, or solitary waves, if they fall off to a constant value away from a guiding center. We shall call the latter
solitons for brevity, even though we have no indication of the integrability of the fluid equations and no multisoliton
solutions.

\subsection{Constant profile solutions}

The conditions for a constant profile configuration moving at speed $u$ are
\be
\rho(x,t) = \rho(x-ut) ~,~~~ v(x,t) = v(x-ut)
\ee
so on any fluid function $\partial_t = -u \partial$. The continuity equation can be solved to give $v$ as a function of $\rho$
\be
v = {c \over \rho} + u
\ee
The integration constant $c$ quantifies the underlying current (`drift'), $c=0$ corresponding to all particles in the fluid moving
at the same velocity $u$. Substituting $v$ in the Euler equation and rearranging yields
\be
\left[{uc\over \rho\sigma} - {1\over 2} {\Bigl({\rho^2 \over{\raisebox{2pt}{$\sigma$}}}\Bigr)}'' \left({c\over {\raisebox{2pt}{$\rho$}}}+u\right)^2
+ {c\sigma\over \rho} \left({\rho \over \sigma^2} \right)' \left({c\over {\raisebox{2pt}{$\rho$}}}+u\right) \right] \d \rho
+ \d \left[{D \over \sqrt{\sigma}} \d \left({D \over \sqrt{\sigma}}\d\rho\right)\right] = 0
\label{flu1}\ee
Remarkably, the expression in the first square bracket is an exact second $\rho$-derivative,
\be
{uc\over \rho\sigma} - {1\over 2} {\Bigl({\rho^2 \over{\raisebox{2pt}{$\sigma$}}}\Bigr)}'' \left({c\over {\raisebox{2pt}{$\rho$}}}+u\right)^2+ {c\sigma\over \rho} \left({\rho \over \sigma^2} \right)' \left({c\over {\raisebox{2pt}{$\rho$}}}+u\right)
= -\left[{(c+u \rho)^2 \over 2\sigma} \right]''
\label{Fr}\ee
which allows to integrate equation (\ref{flu1}) to
\be
{D \over \sqrt{\sigma}} \d \left({D \over \sqrt{\sigma}}\d\rho\right) = \left[{(c+u \rho)^2 \over 2\sigma} \right]' + \half K_1
\label{flu2}\ee
for a constant $K_1$.

The above equation can be solved through a hodographic transformation. We define the hodographic variable $\tau$
through the equation
\be
{dx \over d\tau} = {D \over \sqrt{\sigma}} ~,~~~~{\rm thus}~~~~~ {D \over \sqrt{\sigma}}\d = {d \over d\tau} 
\label{the} 
\ee
and equation (\ref{flu2}) is written
\be
{d^2 \rho \over d\tau^2} =   \left[{(c+u \rho)^2 \over 2\sigma} \right]' + \half K_1
\label{dht}
\ee
A further integral of the above equation is then obtained as
\be
\left({d \rho \over d\tau}\right)^2 =  {(c+u \rho)^2 \over \sigma} + K_1 \rho + K_2
\label{dhV}\ee
with a new constant $K_2$, in a `hodographic energy conservation' expression. Finally, combining (\ref{dhV}) and
(\ref{the}) we can eliminate the hodographic variable and obtain
\be
(\d \rho)^2 = {(c+u \rho)^2 + ( K_1 \rho + K_2 ) \sigma \over D^2} \equiv -W(\rho)
\ee
This is a separable equation, the right hand side representing an``effective potential'' $W(\rho)$,
and is solved implicitly by
\be
\int{D\, d\rho \over \sqrt{{(c+u \rho)^2} + ( K_1 \rho + K_2 )\, \sigma}} = x - x_o
\label{rxsol}\ee
We have in total four integration constants, $c$, $K_1$, $K_2$ and $x_o$, the last one corresponding to translation
invariance and eliminated by an appropriate choice of origin.

\subsection{Remarks on the solutions}

Equation (\ref{rxsol}) suggests that the profile $\rho(x)$ can be considered as the motion of a one-dimensional particle
with coordinate $\rho$ in potential $W(\rho)$ and energy $E=0$, with $x$ playing the role of the evolution parameter.
Finding the solution involves solving the nonlinear differential equation (\ref{rxsol}) and inverting it to determine $\rho(x)$.
This will depend on the specific choices of $D(\rho)$ and $\sigma(\rho)$. Further, whether the solutions are physically
acceptable, with $\rho(x)$ remaining finite and positive everywhere and respecting any other constraints, such as
$\rho \le 1$ in the SEP case, depends on the choice of integration constants.

We can still draw some important
conclusions from the general form of the solution:

{\it i.} For $K_1=K_2 =0$ the solution depends only on the diffusion function $D(\sigma)$, which is 1 in most cases
of interest. Setting $D=1$ and $\exp(-x_o /u)=u\rho_o$ we obtain
\be
\rho = -{c \over u} + \rho_o \, e^{u x} ~,~~~{\rm so} ~~~~ \rho(x,t) = -{c \over u}+ \rho_o \, e^{u (x - u t)}
\ee
Therefore, all fluids share a common set of (unphysical) exponential solutions.

{\it ii.} Static solutions, for which $u=0$ (stationary solitons) and $c=0$ (no drift) depend only on the function
$D(\rho)^2 /\sigma(\rho)$. By contrast, stationary solutions (only $u=0$) and drift-free solutions (only $c=0$) depend
on both $D(\rho)$ and $\sigma(\rho)$.

{\it iii.} For $D=1$ and $\sigma(\rho)$ a polynomial of degree $n$ in $\rho$, the effective potential is a polynomial of
degree $n$ (or 2, if $n<2$), so the form of the solution is generically the same for all such fluids. However, as
will become apparent, the specific form of the effective potential, and in particular its zeros, will be 
crucial for the existence of localized solitons or waves.

We conclude by remarking that traveling wave solutions could be examined in the original formulation of the systems
in terms of $\r$ and $p$ rather than the hydrodynamic formulation in terms of $\r$ and $v$. This was done in \cite{ZaMe}
for the specific case of the KMP fluid, and the results found there are in agreement with the results that we derive in section 
{\bf 5.2}. The method used in \cite{ZaMe} relies on the  $\r,p$ traveling wave equation being of a Hamiltonian form
and it could be generalized to arbitrary $\sigma (\r )$ but requires $D(\rho)$ to be a constant. The use of the hodographic
variable $\tau$ in our derivation can be traced to that fact and circumvents this restriction.

\section{Soliton and wave solutions for specific fluids}

Our main focus is the identification of localized soliton and nonlinear wave solutions in the special fluids of interest, 
and we treat each case in detail.

\subsection{Independent particles}

For independent particles, $D=1$ and $\sigma=2\rho$, the effective potential is quadratic in $\rho$ and $\rho(x)$
corresponds to the motion of a harmonic oscillator, stable or unstable depending on the value of $K_1$. This is
consistent with the solutions of the linear (anti)diffusion equations in terms of $\Phi,\Psi$ and it is easier to work
with these fields. In the present case the (anti)diffusive fields are
\be
\Phi(x,t) = \phi(x-ut) \, e^{-at} ~,~~ \Psi(x,t) = \psi(x-ut) \, e^{at} 
\ee
with the functions $\phi(x)$ and $\psi(x)$ satisfying
\be
\d^2 \phi+u\, \d\phi+a \, \phi = \d^2 \psi - u\, \d \psi + a\, \psi = 0
\ee
The general solution is
\be
\phi = e^{-ux/2} \left(A e^{qx/2} + B e^{-qx/2} \right) ~,~~~
\psi = e^{ux/2} \left(\At e^{qx/2} + \Bt e^{-qx/2} \right) ~;~~~ q^2 = u^2 -4a
\ee
from which
\bea
\rho &=& \phi \psi = A\Bt+\At B+A\At \, e^{qx} + B\Bt \, e^{-qx}\label{inden}\\
v &=& 2\d\theta = {\d\psi \over \psi} - {\d\phi \over \phi} = u + \r {\At B-A\Bt \over \rho}
\eea
The constants $A,B,\At,\Bt,a$ play the role of $c,K_1,K_2,x_o$ in the effective potential solution,
noting that the transformation $A \to \lambda A, B \to \lambda B, \At \to \lambda^{-1} \At, \Bt \to \lambda^{-1} \Bt$
leaves the solution invariant, therefore reducing the number of relevant constants to four.

The above density (\ref{inden}) has a constant part and a sinusoidal or exponential part, depending on the sign of 
$q^2$, and by varying the parameter $a$ (which is invisible at the level of fluid functions $\rho$ and $v$) we can obtain
the full range of solutions for any $\r$. Note, also, that for $q\neq 0$ and $\At B \neq A\Bt$ there is a drift,
even for $u=0$, unlike the static case.

However, the exponential solutions diverge at infinity, while it can be checked that the oscillatory ones become
negative in intervals. Only in the case $A=B$, $\At=\Bt$, we have a positive density repeating periodically between
vanishing points, with a velocity $v=u$ and no drift, which is the Galilean boost of a static solution.
Altogether, these solutions are not very physical.

\begin{figure}
\hskip .5cm
\includegraphics[scale=0.8]{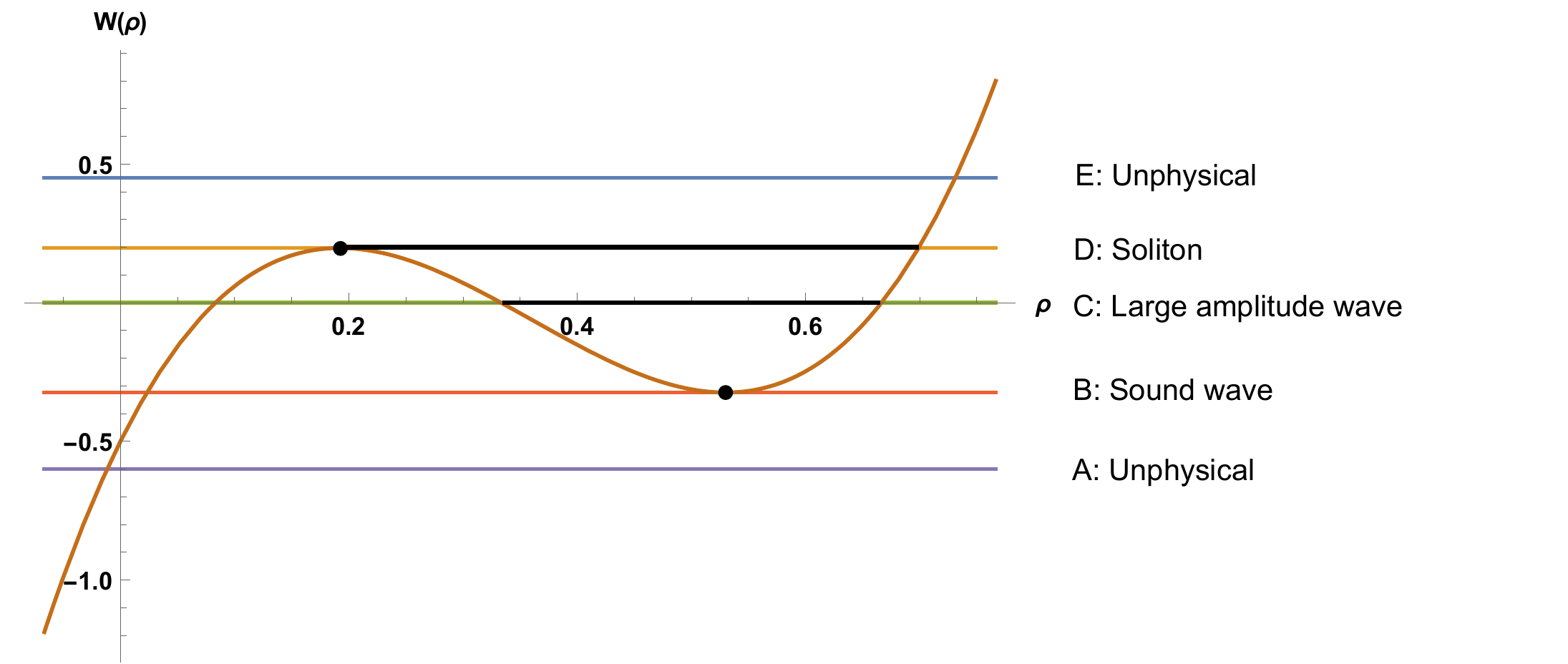}
\vskip -1cm
\caption{Possible solutions for $k_1 >0$}\label{figure1}
\end{figure}

\subsection{KMP fluid}

For the KMP fluid, $D=1$ and $\sigma = 2\rho^2$. The effective potential $W(\rho)$ entering equation
(\ref{rxsol}) is
\be
W = -(c+u \rho)^2 - 2 \rho^2 (K_1 + K_2 \rho) = k_1 \,\rho^3 + k_2 \,\rho^2 - 2c u \,\rho - c^2
\label{WKMP}
\ee
where $k_1 = -K_1$, $k_2 = -2K_2 - 2c u$ are new constants that span the full range of real values.
$W(\rho)$ is a cubic function that behaves generically as in fig.1 or fig.2, depending on the sign of $k_1$ and
the value of the remaining parameters.

Solutions for $\rho(x)$ correspond to motion along the range of $\rho$
where $W\le 0$. Solutions where $\rho <0$ in any interval are excluded as unphysical. Similarly, solutions where
$\rho$ becomes infinite in any point, although mathematically interesting, are also excluded.

The above requirements imply that only potentials $W(\rho)$ that develop a finite well, between a local maximum and a
local minimum, can support physical solutions. Note, also, that $W(0) = -c^2 \le 0$. This leaves the following possibilities:

{\it i.} For $k_1 >0$, the potential must be as in fig.1. Depending on where the line $W=0$ cuts the graph we see that
only cases B, C and D correspond to physical solutions: B to a single constant density configuration with small-amplitude
sound waves; C to a finite amplitude wave,
where $\rho$ bounces periodically between a maximum and a minimum; and D to a soliton, $\rho$ bouncing off
the maximum and going asymptotically to the minimum as $x \to \pm \infty$.

{\it ii.} For $k_1 <0$, the potential must be as in fig.2. In general, cases B, C and D could yield physical solutions.
However, they all require $W(0)=0$, that is, $c=0$, to ensure positivity of $\rho$. For the KMP potential this implies that the linear term $-2c u x$ also vanishes, and $x=0$ is one of the extremal points. Therefore, there are no physical solutions
for the KMP fluid.

{\it iii.} For $k_1 =0$, we need again $c=0$, which also eliminates the linear term leaving only the trivial
vanishing solution with no sound waves. Finally, for $k_1 = k_2 =0$ we obtain either unphysical ($c u \neq 0$) or trivially constant ($c=0$) solutions with no sound waves.

\begin{figure}\vskip -1cm
\hskip 0.8cm
\includegraphics[scale=0.8]{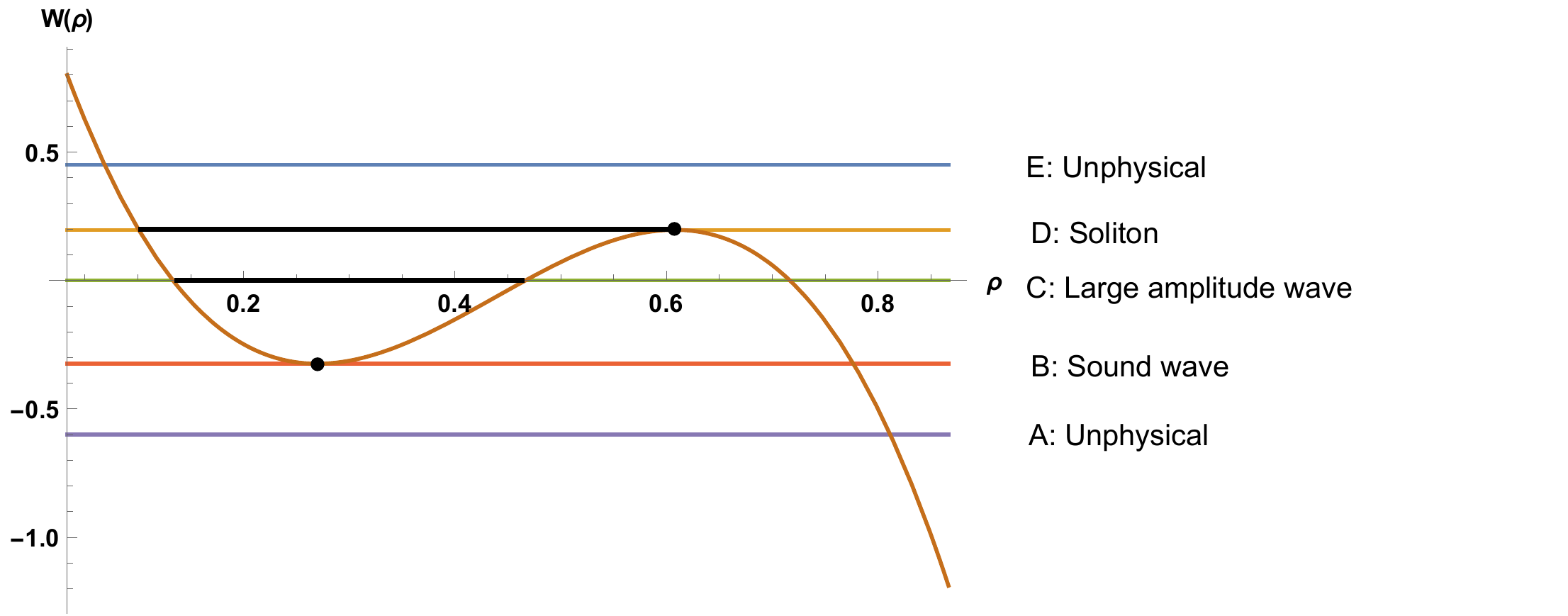}\label{figure2}
\vskip -0.5cm
\caption{Possible solutions for $k_1 <0$}
\end{figure}

This leaves $k_1 >0$ and $W(\rho)$ having three real positive roots as the domain with physical solutions for the KMP fluid:
\be
W (\rho) = k_1 (\rho-\rho_1)(\rho-\rho_2)(\rho-\rho_3)~,~~~ k_1 >0~,~~~ 0 \le \rho_1 \le \rho_2 \le \rho_3
\ee
The three roots $\rho_{1,2,3}$ are related to the parameters $k_2,c,u$ as implied by the form of the potential
in (\ref{WKMP}); that is,
\be
k_2 = -k_1 (\rho_1 + \rho_2 + \rho_3 )~,~~
-2 c u = k_1 ( \rho_1 \rho_2 + \rho_2 \rho_3 + \rho_3 \rho_1 ) ~,~~
c^2 = k_1 \rho_1 \rho_2 \rho_3 
\ee
For $\rho_1 < \rho_2 < \rho_3$ the solution is a finite amplitude wave; for $\rho_1 \to \rho_2$ the solution is a soliton; and for $\rho_2 \to \rho_3$ the solution is a small-amplitude (sound) wave.
We examine the three types of solutions below.

{\bf {5.2.1:}}\, {\underline {\bf Solitons}} are perhaps the most interesting solutions.
In this case the potential $W(\rho)$ is of the form
\be
W_s = k_1 (\rho-\rho_o )^2\, (\rho- \rho_p)
\ee
with $\rho_o$ the background density and $\rho_p$ the density at the peak of the soliton. Comparing with the form 
of $W$ in (\ref{WKMP}) we deduce
\be
k_2 = - k_1 (2\rho_o + \rho_p ) ~,~~~ c = \pm \rho_o \sqrt{k_1 \rho_p} ~,~~~
u = \mp \sqrt{k_1 \over \rho_p} \left({\rho_o \over 2} + \rho_p \right)
\ee
Equation (\ref{rxsol}) in this case can be solved and inverted explicitly to find $\rho(x)$. Setting
$\rho_p = \rho_o +A$, with $A$ the amplitude of the soliton, and $k_1 = k^2 / A$, with $k$ a new parameter,
we obtain overall:
\be
\rho_s (x) = \rho_o + {A \over \cosh^2 {k\over 2} x} ~,~~~ v_s (x) = u + {c \over \rho_s (x)} \label{sv}
 \ee
 \be
\hskip -0.5cm{\rm with}~~~ u = {1+{3\rho_o\over 2A} \over \sqrt{1+{\rho_o \over A}}} \,k ~~,~~~~~
c = -\sqrt{1+{\rho_o \over A}}\,k\, {\rho_o} \nonumber
\ee
A plot of the density and velocity of the soliton is given in fig.3.
\begin{figure}
\vskip -1.6cm
\hskip 2.5cm\includegraphics[scale=0.7]{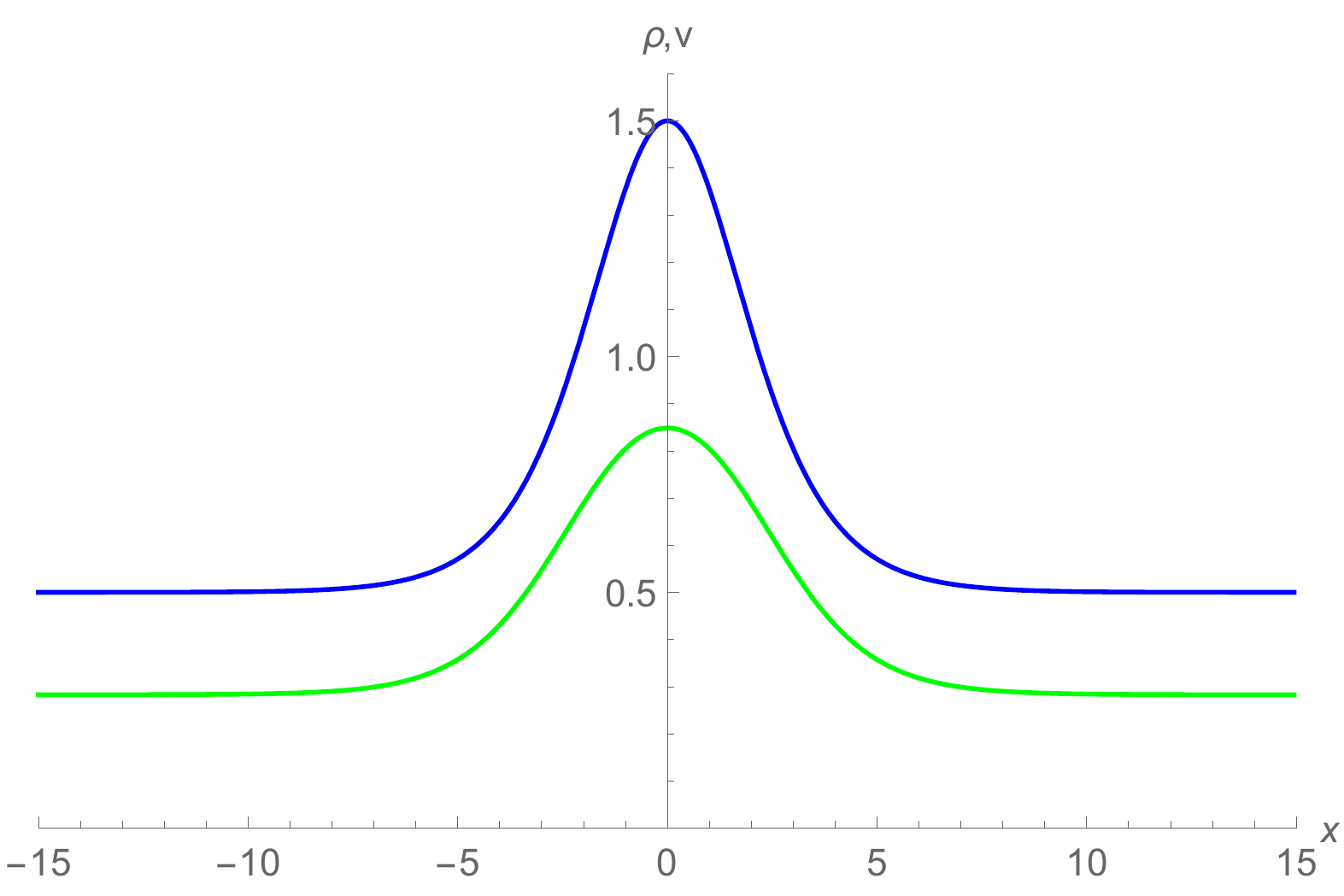}\label{figure3}
\caption{{\bf{\color{blue}Density}} and {{\bf\color{mygreen}velocity}} profiles of the KMP soliton}
\end{figure}

The excess particles carried by the soliton, on top of the background, is
\be
N_s = {4 A \over k}
\ee
Note that the background fluid far away from the soliton has a nonzero velocity
\be
v_s (\pm \infty) = {k\rho_o \over 2A\sqrt{1+{\rho_o \over A}}}
\ee
which is not possible to eliminate, as the fluid is not boost invariant.

The above solution is valid as long as $\rho_o >0$, such that $c \neq 0$. For $\rho_o = c = 0$ the
form of the solution in (\ref{sv}) remains valid, but the equation for $u$ need not be satisfied, and we have
solutions for arbitrary velocity $u$, including $u=0$. Clearly the limit $\rho_o \to 0$ is discontinuous.

{\bf {5.2.2:}} \, {\underline {\bf Finite amplitude waves}} correspond to a potential
\be
W_w = k_1 (\rho-\rho_b)(\rho-\rho_t)(\rho-\rho_c)
\label{wav}\ee
where $\rho_t < \rho_c$ are the trough and crest values of the density, and the positive parameters $k_1$ and
$\rho_b$($<\rho_t$) control the periodicity (wavelength) and speed of the wave. Equation (\ref{rxsol}) now is solved in
terms of elliptic functions, and we will not give their explicit form. We simply state the expression of the wavelength
$\lambda$ of the wave in terms of $u,\rho_t,\rho_c$ and the parameter $\rho_b$ after eliminating $k_1$:
\be
\lambda = 2 {\rho_b (\rho_t + \rho_c ) + \rho_t \rho_c \over u \sqrt{\rho_b \rho_t \rho_c (\rho_c - \rho_b)}} 
 \; K \left({\rho_c - \rho_t \over \rho_c - \rho_b} \right)
\ee
where $K(\,\cdot\,)$ is the K-elliptic function. This is a so-called ``cnoidal'' wave. A numerical plot of the density
and velocity of the wave is given in fig.4.
\begin{figure}
\vskip -1.8cm
\hskip 1.5cm\includegraphics[scale=0.72]{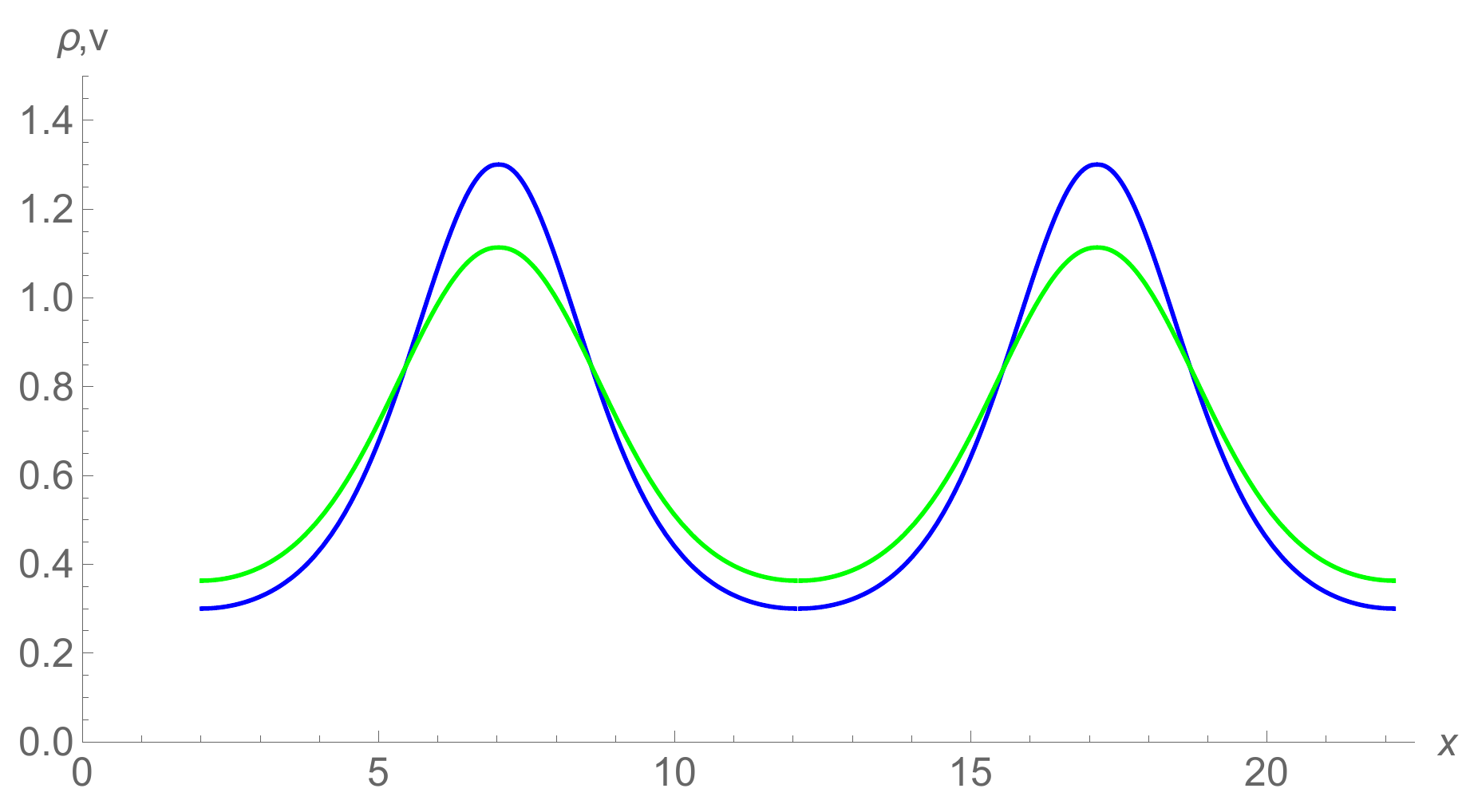}\label{figure4}
\vskip -0.2cm
\caption{{{\bf\color{blue}Density}} and {{\bf\color{mygreen}velocity}} profiles of a KMP nonlinear wave}
\end{figure}

{\bf {5.2.3:}} \, {\underline {\bf Sound waves}} correspond to small-amplitude periodic waves. We could examine them by expanding
the action (\ref{fl}) around a constant background solution, but we can instead recover them by considering the potential
\be
W = k_1 (\rho- \rho_b)\,[ (\rho-\rho_o )^2 - \epsilon^2]
\ee
where $\epsilon \ll \rho_o$ represents the amplitude of the wave. The wavenumber $k$ of the wave is given by the
curvature of the potential at $\rho = \rho_o$
\be
k^2 = \half{W'' (\rho_o )} = k_1 (\rho_o - \rho_b )
\ee
The relation of parameters $k_1,\rho_b$ with $u$ and $c$ are similar to the soliton case. The fluid has a background
density $\rho_o$ and a background velocity
\be
v_o = u + {c \over \rho_o} = u {\rho_o \over \rho_o + 2 \rho_b} = {k \rho_o \over 2\sqrt{\rho_b (\rho_o - \rho_b )}}
\ee
Setting $u= \omega/k$ the phase velocity of the wave, with $\omega$ its cyclic frequency, and eliminating $k_1$ and
$\rho_b$ in favor of $u$ and $v_o$ we obtain the dispersion relation
\be
\omega = k\left(2v_o \pm \sqrt{v_o^2 - {k^2}}\right)
\label{bif}
\ee
We observe that sound waves around a static background
do not exist, but a constant background velocity allows for sound waves of wavenumber
\be
|k| \le v_o
\label{vmin}\ee
and that there is birefringence, with two possible velocities for each wavelength. For long wavelengths the
two speeds of sound are $v_o$ and $3 v_o$.

The existence of sound waves for the KMP hydrodynamic equations is related to the {\bf {stability}} of the system.
For a KMP system on a ring of unit length, wave configurations would have a wavelength quantized to $\lambda = 1/n$, or
$k = 2\pi n$, for integer $n$. The lowest excitation that the system can have is sound waves at the fundamental frequency
$k = 2\pi$. The existence of such sound waves signifies that the constant profile configuration may be destabilized against
fluctuations. Putting $k=2\pi$ in (\ref{vmin}) above, and expressing it in terms of the current $j = \rho v$, we obtain
\be
j \ge 2\pi \rho_o
\ee
This is precisely the critical current derived in \cite{Inst} required for instability to kick in. 
The bifurcation of the dispersion relation (\ref{bif}) at the critical point, leading to birefringence, is a hallmark of the onset of instabilities,
one branch being stable and the other unstable.

We conclude the discussion of KMP fluids by pointing out that the symmetry transformation (\ref{Ksym}) in terms
of fluid variables becomes
\be
\rho \to \theta (1-\rho \theta )~,~~~ \theta \to {1 \over \theta}
\ee
For a constant profile configuration with $v = 2 \rho \d \theta$, $\theta$ will in general increase linearly with $x$, and
the above transformation produces unphysical solutions. Since their SEP counterparts according to the mapping (\ref{mapKS},\ref{mapSK}) would both be physical, we conclude that some acceptable SEP solutions would map to unphysical
KMP configurations.

\subsection{SEP fluid}

Solutions for SEP fluids could in principle be derived from those of KMP fluids using the mapping between the systems. However, given the limitations in the range of applicability of the mapping pointed out in section {\bf 2.2}, and the related fact
that some constant-profile SEP solutions would not correspond to physical KMP solutions as pointed out at the end of the
previous section, it is more reliable to derive the solutions independently.

The effective potential in this case is
\bea
W &=& -(c+u \rho)^2 - 2 \rho (1-\rho) (K_1 + K_2 \rho)\nonumber \\
 &=& k_1 \,\rho^3 + k_2 \,\rho^2 - (k_1 + k_2 +u^2 +2c u ) \,\rho - c^2
\label{WSEP}
\eea
again with two new constants $k_1,k_2$ that span the full range of real values. Note that the reflection symmetry (\ref{refl})
of the model becomes in terms of fluid variables
\be
\rho \to 1-\rho ~,~~~ v \to -{\rho \over 1-\rho} v
\label{Srefl}
\ee
and for a constant profile solution
\be
u \to -u ~,~~~ c \to -(u+c)
\ee

The analysis of possible solutions is the same as in the KMP case, with some important differences. First,
the density must remain in the range $0 \le \rho \le 1$; and second, when $k_1 <0$ we can choose $c=0$ without
the linear term vanishing, so the case $k_1 <0$ is not eliminated. In fact, we can see that taking $\rho \to 1-\rho$
in (\ref{WSEP}) $k_1 \to -k_1$, so the cases $k_1 >0$ and $k_1 <0$ are related by the reflection symmetry. In particular,
solitons for $k_1 <0$ are actually antisolitons.
So we can in principle examine only one case, and obtain the remaining solutions by applying (\ref{Srefl}), but we prefer to
present them both in parallel for clarity.

As in the KMP case, we obtain physically acceptable solutions when $W(\rho)$ has three nonnegative
(possibly degenerate) zeros. Putting
\be
W = k_1 (\rho-\rho_1)(\rho-\rho_2)(\rho-\rho_3)~,~~~ k_1 >0~,~~~ 0 \le \rho_1 \le \rho_2 \le \rho_3 \le 1
\ee
and comparing with (\ref{WSEP}) we obtain
\be
c^2 = k_1 \rho_1 \rho_2 \rho_3 ~,~~~ (u+c)^2 = -k_1 (1-\rho_1 )(1-\rho_2 )(1-\rho_3 )
\ee
For $k_1 > 0$ the second quantity is nonpositive, and the only acceptable solution is when $\rho_3 =1$, $u=-c$.
So the fluid must necessarily reach $\rho=1$ at some point (or points).
Conversely, for $k_1 <0$ we must have $\rho_1 = 0$ and $c=0$, and the density must vanish at some points.

Overall, we see that we have one less parameter than in the KMP case, as we have no control over either the crest or
the trough value of the density. Otherwise the solutions look similar to the KMP ones.

{\bf {5.3.1:}} \,{\underline {\bf Solitons}} correspond to $k_1 >0$, $\rho_1 = \rho_2 = \rho_o$, $\rho_3 =1$. We obtain
\be
\rho = \rho_o + {1-\rho_o \over \cosh^2 {k x \over 2}} ~,~~~~ v = - k {\rho_o \sqrt{1-\rho_o} \over \rho_o +
\sinh^{-2} {k x \over 2}}
\ee
Their traveling speed $u$ and excess particle number $N_s$ are
\be
u = k {\rho_o \over \sqrt{1-\rho_o}} ~,~~~~ N_s = 4{1-\rho_o \over k} = 4{\rho_o \sqrt{1-\rho_o} \over u}
\ee
We note that the underlying velocity is opposite to their speed, signaling a strong drift, reaching an asymptotic value
of $v_\infty = -u (1-\rho_o )/\rho_o$ away from the soliton.

{\bf {5.3.2:}} \,{\underline {\bf Antisolitons}} correspond to $k_1 <0$, $\rho_1 =0$, $\rho_2 = \rho_3 = \rho_o$ and are related to solitons through the reflection symmetry (\ref{refl}). We obtain
\be
\rho = \rho_o - {\rho_o \over \cosh^2 {k x \over 2}} ~, ~~~ v = u = k {1-\rho_o \over \sqrt{\rho_o}} ~,~~~
N_{as} = -4{\rho_o \over k} = -4{(1-\rho_o ) \sqrt{\rho_o} \over u}
\ee
Interestingly, antisolitons have no drift since $v=u$. (Note that although the current $j=\rho v$ changes sign under
reflection, the velocity $v$ does not.)

At half-filling, $\rho_o = 1/2$, solitons and antisolitons become symmetrical, although not entirely so, since the soliton
still has a drift. Remarkably, if we arrange for the asymptotic soliton fluid velocity $v_\infty$ and antisoliton velocity $u$
to match, such that a well-separated soliton-antisoliton pair be an asymptotic solution, then soliton and antisoliton travel
at opposite speeds, have the same profile and carry equal and opposite particle number, suggesting a particle-antiparticle
scattering event.

{\bf {5.3.3:}} \,{\underline {\bf Finite amplitude waves}} for $k_1 >0$ correspond to $\rho_1 := \rho_b < \rho_2 :=\rho_t$, $\rho_3 =1$,
with $\rho_t$ representing the trough density of the wave (the crest density is necessarily 1). These waves have speed
\be
u = -c = \sqrt{k_1 \rho_b \rho_t}
\ee
As for solitons, there is one less parameter than the KMP case since the crest density is fixed. 
The parameter $\rho_b$ fixes both the wavelength and the speed of the wave and the solution for $\rho$ is given
by an elliptic function. The expressions of the wavelength of the wave in terms of $\rho_b, \rho_t$ and $u$ are
\be
\lambda = {4 \over u} \sqrt{\rho_b \rho_t \over 1-\rho_b}  
 \; K \left({1 - \rho_t \over 1 - \rho_b} \right)
\ee
The dual wave with trough density reaching zero, corresponding to $k_1 <0$, $\rho_1 =0$, $\rho_2 = \rho_c < \rho_3 = \rho_b$,
has a wavelength given by
\be
\lambda = {4 \over u} \sqrt{(1-\rho_b )(1- \rho_c ) \over \rho_b}  
 \; K \left({\rho_c \over \rho_b} \right)
\ee

{\bf {5.3.4:}} \,{\underline {\bf Sound waves}} can exist only over background densities $\rho_o =1$ and $\rho_o =0$. Sound waves at
full filling ($\rho_o = 1$) are actually degenerate as the fluid can have no velocity, since $v = (1-\rho ) \d \theta$ identically
vanishes. Similarly, 
sound waves around its reflection symmetric empty state $\rho_o = 0$ are also essentially degenerate, since the underlying fluid moves at
the velocity of propagation $u$ of the wave, so the entire wave propagation is due to fluid transport. 
Sound waves can propagate at any wavelength for any frequency over these degenerate backgrounds.
Working with
$k_1 <0$, $\rho_1 = \rho_2 = 0$, $\rho_3 = \rho_b$ we obtain $u = \sqrt{-k_1 \rho_b}$ and a linear dispersion
relation for any speed $u$,
\be
\omega = u\, k
\ee

\section{Conclusions and outlook}

The existence of soliton and wave solutions is an interesting property of diffusion process fluids. The soliton profiles,
in particular, are typical of soliton solutions in integrable systems. This raises the possibility that the underlying structure
of these fluids is also integrable. Other fluid mechanical systems, such as the hydrodynamics description
of the Calogero-Sutherland model, admit true solitons as these sytems are integrable \cite{AP,AB}.
However, in the absence of multi-soliton solutions, or an analytic derivation of conserved
quantities or a Lax pair, the integrability of diffusion process fluids remains an open issue.

Although in this paper we studied in detail the SEP and KMP systems, our methods, and in particular equation (\ref{rxsol}),
are completely general and work for any $D(\rho)$ and $\sigma (\rho)$. There are other instances in the literature where
general results can be derived for arbitrary transport coefficients, such as, e.g., \cite{MeKr,GaKr}. In particular, if $D(\rho)=1$ and
$\sigma(\rho)$ is a quadratic polynomial, the effective potential $W(\rho)$ is quartic. For such cases (\ref{rxsol}) has the same generic
structure and is solved in terms of elliptic functions. The fact that quadratic $\sigma(\rho)$ can generically be mapped to
the SEP system by a linear rescaling \cite{DG} is also suggestive. The specific physical results for solitons and waves,
however, seem to be quite different between the SEP and KMP systems. A relevant (and straightforward) exercise would be
to examine the solutions of generic quadratic systems and see if they fall on the SEP or KMP side.

The most interesting question is the implications of the solutions identified in the present work for large fluctuation properties
of the underlying statistical processes. The hydrodynamic variable $v$, in particular, has
no direct interpretation for the diffusion process, where the original variable $p$ is more relevant to the analysis of the
statistics of their large fluctuations. The relation of the emergence of sound waves to the onset of instabilities pointed out in
section {\bf 5.2.3} is consistent with the fact that solitons and waves are related to instabilities. In fact,
the appearance of soliton-like configurations in the numerical work of \cite{HuGa} makes it clear that our solutions are
relevant to the destabilizing fluctuations in the corresponding statistical systems.
For a compact space (ring), where such instabilities manifest, there is no sharp distinction between solitons, nonlinear waves and
sound waves. Further, \cite{Inst} established a link between transitions to an unstable phase and the
breakdown of the ``additivity principle'' \cite{Add}. It would be nice to have a unified description of solitons, instabilities and additivity
in which the results of this paper will be put in context and may be of use. These and other relevant questions on the physical interpretation of soliton
and wave solutions are left for future work.

\vskip 0.6cm

\noindent
{\bf{Acknowledgments}}

\noindent
I am indebted to Paul Krapivsky and Kirone Mallick for illuminating conversations and for many useful remarks and suggestions
that helped make this a better manuscript. Kirone's 2018 ICTS presentation, in particular, was responsible for sparking my
interest in the subject. I also thank the anonymous referees for their feedback and remarks. This research was supported in part by the National Science Foundation under grant NSF 1519449 and
by an ``Aide Investissements d'Avenir'' LabEx PALM grant (ANR-10-LABX-0039-PALM).

\end{document}